\documentclass{ws-p9-75x6-50}

\usepackage[]{epsf}
\usepackage[]{amsmath}
\usepackage[]{latexsym}
\usepackage[]{xspace}

\newcommand\ea{et~al.}
\newcommand\tf{\ensuremath{\tau_{\!f}}\xspace}
\newcommand\n[1]{\ensuremath{n_{#1}}\xspace}
\newcommand\nzero{\n{0}}\newcommand\ns{\n{s}}
\newcommand\ti{\ensuremath{t_{i}}\xspace}

\newcommand\ts[1]{\ensuremath{t'_{#1}}\xspace}
\newcommand\tsi{\ensuremath{t'_{i}}\xspace}

\newcommand\dt{\ensuremath{\delta t}\xspace}
\newcommand\dti{\ensuremath{\delta t_i}\xspace}

\newcommand\secref[1]{Sec.~\ref{#1}}
\newcommand\figref[1]{Fig.~\ref{#1}}
\newcommand\Eqref[1]{Eq.~\eqref{#1}}
\newcommand\pot[1]{\ensuremath{10^{#1}}\xspace}
\newcommand\cpot[1]{\ensuremath{\!\cdot\!10^{#1}}\xspace}
\newcommand\mpot[1]{\ensuremath{\mathrm{m}^{#1}}\xspace}

\newcommand\vsq{\mean{v^2}\xspace}
\newcommand\mean[1]{\ensuremath{\langle #1 \rangle}\xspace}

\newcommand\x{\ensuremath{\mathbf{x}}\xspace}

\newcommand\dispers{\ensuremath{\langle \Delta {x'}^2 \rangle}\xspace}

\newcommand\dm{\ensuremath{D_{m}}\xspace}

\newcommand\tg{\ensuremath{T_{g}}\xspace}
\newcommand\pg{\ensuremath{p_{g}}\xspace}

\newcommand\msol{\ensuremath{\mathrm{M}_{\odot}}}

\newcommand\pict[8]{
\begin{figure}[ht]
 \unitlength=\hsize
   \IfFileExists{#1}{
      \begin{picture}(#4,#5)%       
         \put(#6,#7){
            \epsfxsize= #8 \hsize
            \epsffile{#1}}
      \end{picture}}
      {
      \fbox{
         \begin{picture}(#4,#5)      
            \typeout{eps-file #1 not found! draw frame instead}\typeout{}
         \end{picture}}} 
   \caption[]{#2} \label{#3}   
\end{figure}}

\def\AA{\em A\&A}
\def\ASR{\em Adv. Space Res.}
\def\APJ{\em ApJ}
\def\CPC{\em Comp. Phys. Commun.}
\def\ICA{\em Icarus}
\def\JCP{\em J. Comp. Phys.}
\def\JP{\em J. Phys.}
\def\PR{\em Phys. Rev.}

\begin{document}

\title{N-body calculations of cluster growth in proto-planetary disks}

\author{S. Kempf}

\address{Max-Planck-Institut f{\"u}r Kernphysik, Saupfercheckweg 1,
  69117 Heidelberg, Germany\\E-mail: Sascha.Kempf@mpi-hd.mpg.de}

\author{S. Pfalzner and Th. Henning}

\address{Astrophysikalisches Institut, FSU Jena, Schillerg{\"a\ss}chen
  2-3, 07745 Jena, Germany\\E-mail: pfalzner@astro.uni-jena.de \quad
  henning@astro.uni-jena.de}


\maketitle

\abstracts{ We investigated numerically the dust growth driven by
  Brownian motion in a proto-planetary disc around a solar-type young
  stellar object.  This process is considered as the first stage in
  the transformation of the initially micron-sized solid particles to
  a planetary system.  In contrast to earlier studies the growth was
  investigated at the small particle number densities typical for the
  conditions in a proto-planetary disc. Under such circumstances, the
  mean particle distance exceeds the typical aggregate diameter by
  orders of magnitude, and a collision will be a very rare event. We
  derived a criterion which allows an efficient detection of
  candidates for imminent collisions. The N-particle-method we used is
  based upon an adaptive time step scheme respecting the individual
  dynamical states of the aggregates.  Its basic concept is to perform
  on average constant ``length steps'', instead of using constant time
  steps.  The numerical cost of the algorithm scales with the particle
  number better than $N \log N$.  In order to minimise the influence
  of the decreasing number of particles within the simulation box, a
  new rescaling method is used throughout the aggregation process. Our
  numerical results indicate that at very low number densities, the
  growth process is influenced by spatial number density
  fluctuations.}

%%%%%%%%%%%%%%%%%%%%%%%%%%%%%%%%%%%%%%%%%%%%%%%%%%%%%%%%%%%%%%%%%%%%%%%%%%%%%
%% Introduction
%%%%%%%%%%%%%%%%%%%%%%%%%%%%%%%%%%%%%%%%%%%%%%%%%%%%%%%%%%%%%%%%%%%%%%%%%%%%%
\section{Introduction} \label{sec:intro}

Since the advent of high performance computers, scientists have been
able to numerically investigate complex astronomical phenomena that
have interested astronomers and physicists for centuries. The first
problems in numerical astrophysics were hydrodynamical and radiative
transfer problems. Then, starting in the early 1980s, several workers
began modeling the gravitational clustering of stars and galaxies
efficiently using tree codes\cite{pfalzner:96}. Since the beginning of
the 90' Particle codes are although used to investigate smaller-scale
astrophysical problems, such as the formation of the precursors of
the planets (planetesimals).\\

Nowadays it is a generally accepted view among the astrophysicists,
that the genesis of a planetary system coincides with the formation of
sun-like young stellar objects surrounded by a gaseous disc. The
building bricks of the planetesimals are micron-sized solid particles
(the so-called cosmic dust) embedded in the gas of the disc.  The
formation of a planetary system is a surprisingly fast process. From
both, the astronomical observations of proto-planetary discs and
geophysical studies, one can derive a formation time scale of a few
million years. Up to the mid-90s the majority of the astrophysicists
were convinced that the transition of the micron-sized dust grains to
the kilometre-sized planetesimals were due to gravitational
instabilities\cite{safronov:72,goldreich:73}. So it was quite a
surprise for the community, when
Weidenschilling\cite{weidenschilling:95} and Cuzzi \ea\cite{cuzzi:93}
demonstrated that the inherent gas turbulence prevents dense dust
layers from collapsing into a planetesimal. Another possible process
for forming a planetesimal in a proto-planetary disc is the dust
growth due to collisional sticking - the so-called coagulation.  For
particles to collide and stick, there must be a relative velocity
component between the grains.
%The sticking of the colliding grains is 
%caused by binding surface forces\cite{blum:98}. 
In the onset of grain growth Brownian motion dominates other
motions\cite{blum:96}. However, it can not be the source of an
effective grain growth leading to planetesimals, because the growth
time scales are much longer than the few million years.  Consequently,
there has to be a much more effective mechanism for driving the dust
coagulation. Such a source of relative velocities could be the
turbulent gas motion\cite{voelk:80} if the typical time \tf a dust
aggregate needs to couple with the gas stream, is larger than the
typical life time of the smallest turbulent
eddy\cite{weidenschilling:84,mizuno:88}. If one considers realistic
values for a proto-planetary disc around a solar-like
protostar\cite{bell:97}, then one finds that the \tf of the initial
particles must increase by five orders of magnitude in order to have
grain growth driven by turbulence\cite{kempf:98a,kempf:98b}.
Consequently, during the Brownian stage of dust growth the typical
friction time of the dust aggregates has to grow rapidly.  However,
\textit{Monte-Carlo} type simulations of the cluster
growth\cite{meakin:84,brown:85,sablotny:95} indicate that the growth
tends to form fractal aggregates characterised by an almost constant
coupling time \tf. In these simulations, particles are added to the
growing dust grain according to some given rule. The disadvantage of
such Monte-Carlo methods is that they require significant
simplification of the astrophysical conditions. In addition, the Monte
Carlo approach cannot address the inherently dynamic nature of the
coagulation process.\\

To overcome the disadvantages of Monte-Carlo astrophysical
simulations, we developed a self-consistent N-particle code. In order
to obtain a detailed understanding of the coagulation in
proto-planetary discs, we considered the influence of the aggregate
structure in the coagulation process. The ``classical'' tool for a
numerical study of this phenomenon is the molecular dynamic (MD).
However, since the dust grains are performing a non-deterministic,
diffusive motion, it is not straight-forward to apply the MD concept
for modeling this process.  Another difficulty with the MD approach is
that we wish to study the coagulation at very low particle number
densities. Or more precisely, the mean particle distance $\mean{\x} =
\{3/4 \pi n\}^{1/3}$ ($n$ - particle number density) exceeds the
typical aggregate size by at least four orders of magnitude. Under
such circumstances, a mutual collision of two particles is a rare
event. In order to model the coagulation exactly, it is essential that
no collisions are missed. To meet this requirement one needs a good
estimate for the particle position after a certain period of time.  In
the following we demonstrate that, for sufficient small time steps, we
can use the MD methods even for diffusing particles.

%%%%%%%%%%%%%%%%%%%%%%%%%%%%%%%%%%%%%%%%%%%%%%%%%%%%%%%%%%%%%%%%%%%%%%%%%%%
%
% computational technique
%
%%%%%%%%%%%%%%%%%%%%%%%%%%%%%%%%%%%%%%%%%%%%%%%%%%%%%%%%%%%%%%%%%%%%%%%%%%%
\section{Computational technique \label{sec:comp}}

\subsection{Simulation of diffusive trajectories \label{ss:equ_mot}}

Since the motion of growing dust grains is only affected by the
gas-grain interaction (diffusion), and in some cases by the
gravitational field, the particle dynamics between successive
collisions can be treated independently. The ``equation of motion''
for diffusive particles is given by the Langevin equation
\begin{eqnarray}
  \mathrm{d}_t x &=& v \nonumber \\
  \mathrm{d}_t v &=& -\tf^{\!-1} v + m^{-1} F_{fl}.
  \label{le1}   
\end{eqnarray}
Here, $x$ and $v$ are the Cartesian components of the position and
velocity of the particle. Furthermore, the friction time \tf is the
mean interval the particle needs to dissipate its relative kinetic
energy with respect to the suspending gas. The influence of the
gas-grain interaction is approximated by the stochastic force $F_{fl}$
having the properties $\mean{F_{fl}}=0$ and $\mean{F^{2}_{\!fl}} =
2mkT/\tf$ ($m$ - particle mass, $T$ - kinetic
gas temperature, $k$ - Boltzmann constant).\\

The diffusive path of the particle can be obtained by successive
integration of \eqref{le1}.  After a time interval $\tau=\delta t
/\tf$, the new particle position and velocity are given
by\cite{chandra:43,ermak:80}
\begin{eqnarray}
  x&=&x_0 +\tf(v\! + \! v_0) \tanh {\left( \tau/2 \right) } + 
   \left\{2 \tf^{2} \mean{v^2}
                 [\tau-2 \tanh{\left( \tau/2 \right)}]
                \right\}^{1/2} \xi_1 \label{x_sto} \\
  v&=&v_0 e^{-\tau} + \left\{\mean{v^2}  [1-e^{-2 \tau}]\right\}^{1/2} \xi_2,
  \label{v_sto}
\end{eqnarray}
where $\xi_1$ and $\xi_2$ are normalised Gaussian random numbers, i.e.
$\mean{\xi_i}=0$, $\mean{\xi_1 \xi_2}=0$, and $\mean{\xi^{\!2}_i}=1$
($i=1,2$).  As mentioned before, the MD concept requires knowledge of
the particle position after a period of time. For a particle following
a diffusive trajectory, it is impossible to obtain such an expression
for an arbitrary interval $\delta t$, since only the probability of
the particle being at position $x$ after $\delta t$ can be derived.
However, on time scales $\delta t\!\ll\! \tf$, the particle dynamics
is still strongly correlated with its initial state.  Performing a
first order series expansion of \Eqref{x_sto} and \eqref{v_sto} yields
\begin{eqnarray}  
  x&=&x_0 + v_0 \delta t + 
  \mathcal{O}(\tau^{3/2})
  \label{x_bal} \\
  v&=&v_0 \left[1-\tau \right] +  \sqrt{2 \mean{v^2}\tau} \xi + 
  \mathcal{O}(\tau^{3/2}).
  \label{v_bal}
\end{eqnarray}
Although \Eqref{x_bal} describes the motion of a free particle
following a ballistic trajectory, the particle is moving
stochastically for $\tau \ll 1$. Due to the ``stochastic force'' $\xi$
in \Eqref{v_bal}, the motion is not reversible in time.  For this
reason, stochastic path segments described by \Eqref{x_bal} and
\eqref{v_bal} are called {\it pseudo ballistic}.\\

To obtain a test criterion for particle collisions, we make use of the
fact that on time scales $\tau \ll 1$ the deviation of the actual
trajectory from the ballistic path is small. From the higher terms of
\Eqref{x_bal}, one finds, that on average the dispersion of the
particle position is
\begin{equation}
    \mean{(x-x_0-v_0 \delta t)^2} = \dispers =
    \tfrac{2}{3} \vsq \tf^2 \tau^3 + \mathcal{O}(\tau^4).
    \label{eq:x_stoch}
\end{equation}
The dispersion in the direction of the ballistic trajectory
\eqref{x_bal} is negligible, while perpendicular to the trajectory,
\dispers contributes significantly. Heuristically, this result can be
interpreted that the particle is moving towards its ballistic path
\eqref{x_bal} within a cone having the time-dependent opening radius
$\dispers^{1/2}$.
\pict{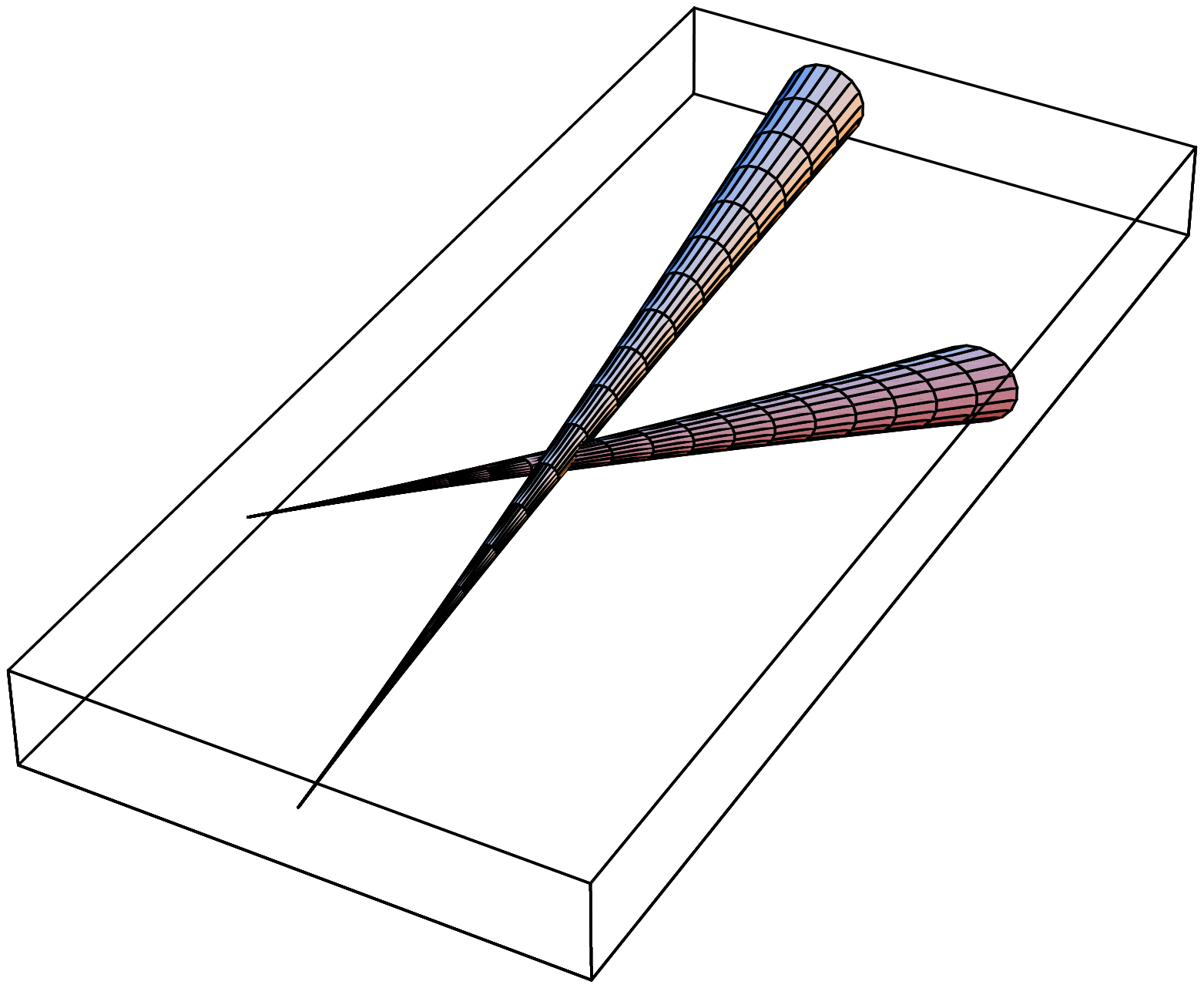} {Example for two particles performing a
  pseudo-ballistic motion. A mutual collision can only happen with
  significant probability within the intersection of the cones.}
{pic:coll_crit}{.95}{0.375}{0.2}{-0.16}{0.5}
Consequently, the prerequisite for an aggregate collision within $\tau
\ll 1$ is the intersection of their probability cones\cite{kempf:98a}.
If a particle pair fulfills this criterion, the trajectories of the
two aggregates will be simulated with a high temporal resolution
according to \Eqref{x_bal} and \eqref{v_bal}. the time step interval
is obtained by the requirement that the faster particle move one
diameter of the larger aggregate during that time. If the particles
become closer to each other than the distance of the sum of their
radii, then they will be combined into a new cluster.

%%%%%%%%%%%%%%%%%%%%%%%%%%%%%%%%%%%%%%%%%%%%%%%%%%%%%%%%%%%%%%%%%%%%%%%%%%%
%
% calendar algorithm
%
%%%%%%%%%%%%%%%%%%%%%%%%%%%%%%%%%%%%%%%%%%%%%%%%%%%%%%%%%%%%%%%%%%%%%%%%%%%
\subsection{Numerical realisation: Calendar algorithm}

As mentioned before the coagulation of particles following a diffusive
trajectory can be represented by MD methods if the applied time step
$\delta t$ is much smaller than the friction time \tf. In the course
of a time step $\delta t \ll \tf$, only particles closer than a
critical separation are likely to collide. Only for these pairs the
test of the collision criterion is required. Therefore, at each time
step the nearest neighbour of the considered particle has to be
determined. Experience shows that, due to the large velocity
dispersion of the growing ensemble, algorithms with global time step
schemes are inefficient for simulating the coagulation. Therefore we
utilised a time step scheme respecting the dynamical state of the
particles.  The algorithm is similar to MD methods simulating hard
sphere fluids\cite{rapaport:95}. The basic idea is to perform average
constant length steps, instead of constant time steps.  To implement
this idea, the simulation volume is divided into cubical cells of the
same size. Next, the individual time step $\dti$ for a particle is
defined by the interval $\dti=\tsi-\ti$ that the particle needs to
move from its current cell to a neighbouring one, where \ti and \tsi
are the instant of the last update of its coordinates (its ``local
time'') and the instant of crossing the cell boundary, respectively.
Obviously, within the interval $(\ti \ldots \tsi)$, a particle can
only collide with particles also belonging to its cell. It is
essential that for the ensemble particle $i$ leaving its cell first
(ie \tsi = $\min(\ts{1} \ldots \ts{N})$), all possible collision
partners have moved into its cell before \ti. This leads to the
following procedure: First, for each cell, build a list of the
embedded particles.  Next, for the particle with the smallest cell
crossing time \tsi, the collision criterion is checked with respect to
the other particles in its cell. If no collision is found, the
particle position and velocity are updated according to \Eqref{x_bal}
and \eqref{v_bal}.  Finally, the cell list is updated and the
procedure is repeated for the next particle with the momentarily
smallest \tsi (see \figref{pic:cell_example}). It should be noted that
in contrast to ``classical'' MD, touching
\pict{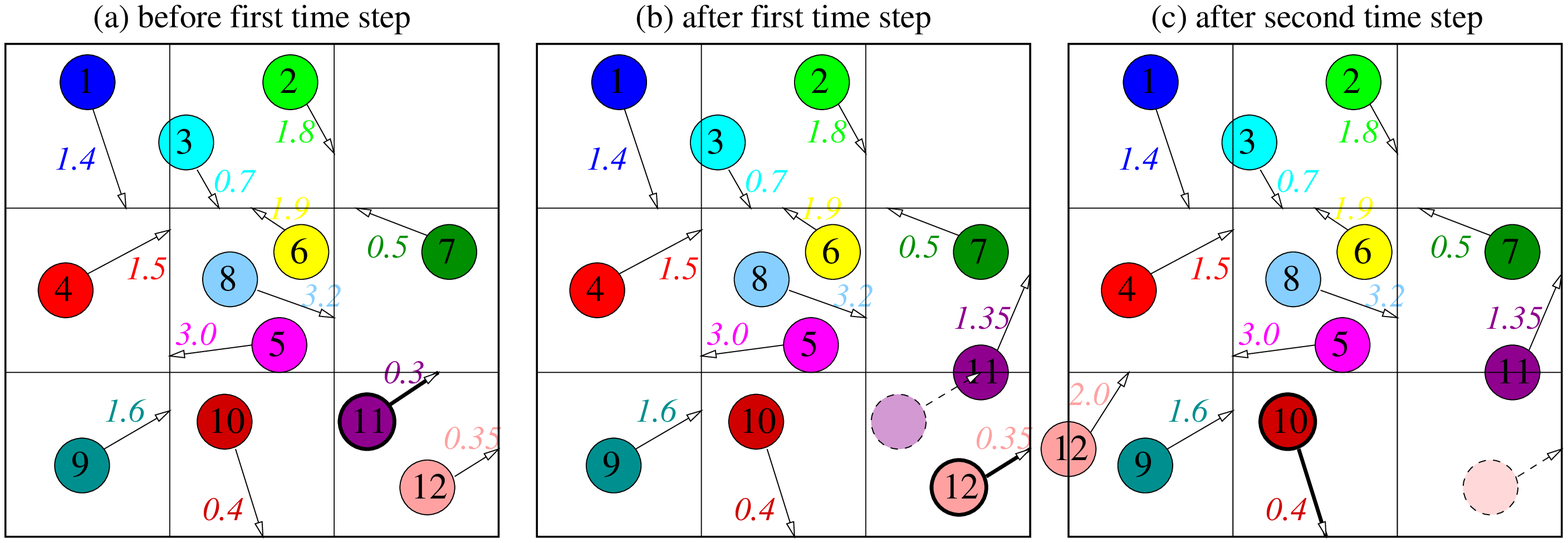}{A 2-dimensional example of three time steps
  according to the presented event-driven algorithm. The arrows show
  the predicted trajectories within the cells, while the italic
  numbers are the predicted instants \tsi of crossing the cell
  boundaries.  For each time step the trajectory of the bold-printed
  particle is simulated. Before the first time step all particles have
  the local time $\ti=0$ (Fig. a). Due to $\dt_{11}=\min(\ts{1} \ldots
  \ts{12})=0.3$, the first time step has to be conducted for particle
  11.  After this, the new local time for particle 11 is $t_{11}=0.30$
  (Fig. b), while after the second time step ($\dt_{12}=\min(\ts{1}
  \ldots \ts{12})=0.35$) the local time for particle 12 is
  $t_{12}=0.35$ (Fig. c). Note that for particle 12 periodic boundary
  conditions were used.}{pic:cell_example}{.95}{0.32}{0.01}{0.}{.95}
the cell boundary at \tsi does not necessarily mean that the particle
will instantaneously leave its embedding cell. Due to the stochastic
nature of its motion, the particle can be ``reflected'' on the
boundary.  For a more detailed consideration of such effects we refer
to the description of our method\cite{kempf:99}.\\

\pict{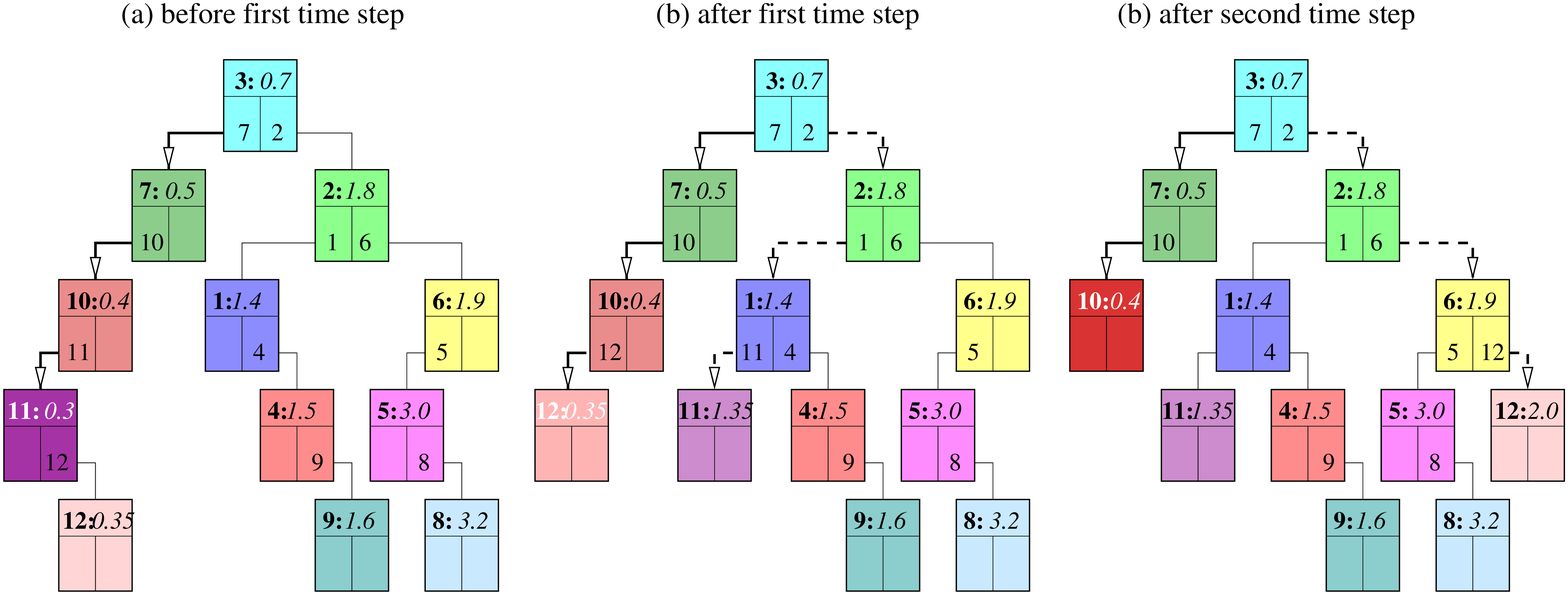} {Evolution of the event tree of the particle
  ensemble illustrated in \figref{pic:cell_example}. The node colours
  correspond with the particle colours in \figref{pic:cell_example}.
  In the upper part of the node the node index (identical with the
  particle index) as well as the predicted instants of cell crossing
  are given.  The lower left number is the index of a linked later
  event, while the lower right number is the index of a linked earlier
  event.  The bold arrows mark the search path to the earliest event,
  while the broken arrows identify the search path for sorting in a
  new event.} {pic:tree_example}{.95}{0.37}{0.01}{0.}{.975}

The prerequisite for an efficient numerical implementation of the
described procedure is a fast search algorithm for the smallest cell
crossing time \tsi (the so-called ``event''). Therefore, the
cell-crossing events are arranged in a tree-like structure (the
so-called ``calendar'') as demonstrated in \figref{pic:tree_example}.

\subsection{Rescaling of the particle number}

Every N-particle method suffers from finite size effects due to the
restricted number of simulated particles $N$. In order to minimise
such effects, one chooses $N$ as large as possible. However, for
aggregation studies it is not sufficient just to start with a large
$N$, since the particle number density decreases due to the
collisional sticking. In previous aggregation studies, the growth was
simulated in a box of constant volume. In that case, the total number
$N$ of simulated particles in the course of the simulation decreased,
leading to strong finite size effects. In extreme cases, the box
contained eventually only a single huge particle (eg
Ziff\cite{ziff:84}), which is an unphysical result. Such effects can
be avoided if the number of simulated aggregates is kept constant on
average.  \pict{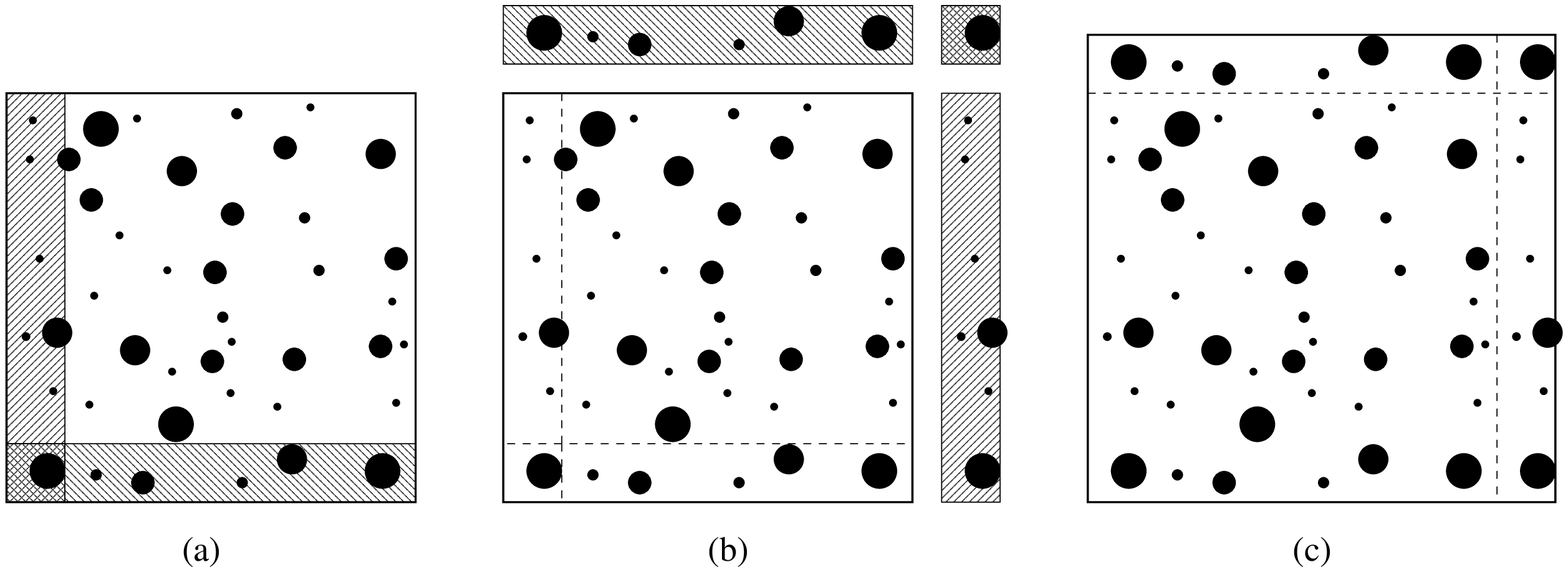} {Two-dimensional example for the
  ``refreshment'' of the particle loss $\Delta N=N(t)-\mean{N}$ by
  rescaling the box size. Fig.~(a) shows the box before the rescaling.
  The hatched parts mark the areas copied to the new volume as
  demonstrated in Fig.~(b). On the average the rescaled box contains
  again \mean{N}~particles (Fig.~(c)).}
{fig:rescale}{0.95}{0.37}{-0.025}{0.}{1.}  For maintaining an almost
constant \mean{N} we rescale the size $L$ of the box according to $L
\rightarrow L'=L \sqrt[3]{1+\Delta N(t)/\mean{N}}$ if the particle
loss $\Delta N$ exceeds a given threshold. The new volume is refilled
with copies of clusters of the original box (\figref{fig:rescale}).
Note, that the global observables of the simulation (mass density,
cluster size spectrum) are not changed by this rescaling method.

%%%%%%%%%%%%%%%%%%%%%%%%%%%%%%%%%%%%%%%%%%%%%%%%%%%%%%%%%%%%%%%%%%%%%%%%%%%
%
% results
%
%%%%%%%%%%%%%%%%%%%%%%%%%%%%%%%%%%%%%%%%%%%%%%%%%%%%%%%%%%%%%%%%%%%%%%%%%%%
\section{Results \label{sec:results}}

In this section we give a brief overview of the results we obtained
from simulations using scalar machines\cite{kempf:98a,kempf:98b}. We
investigated the coagulation driven by Brownian motion for the typical
conditions in a proto-planetary disc at 10 AU\footnote{ We considered
  a disc of 80 AU with a mass accretion of
  $\dot{M}=\pot{-6}\msol\mathrm{yr}^{-1}$ and an ``alpha''-viscosity
  of $\alpha= \pot{-2}$. Using the one-dimensional hydrodynamical disc
  model of Ruden\&Pollack\cite{ruden:91} we determined the gas
  temperature \tg and pressure \pg at 10 AU as about 100 K and 2
  \cpot{-4}Pa, respectively.}. All simulation were started with $N$
randomly distributed spheres of 1 micron diameter. The simplification
of an initially mono-disperse size distribution is justified by the
fact that the growth driven by Brownian motion tends to form
aggregates of similar size. In order to study the influence of the
initial dust number density \nzero simulations with \nzero between
$\pot{7}\ldots\pot{9} \mpot{-3}$ were performed. Most of the
simulations were carried out with $10\,000$ particles. For each
parameter set at least six runs with different seeds of the random
generator were executed.

\subsection{Structure of the growing aggregates}

As outlined in \secref{sec:intro} the prerequisite for an effective
grain growth driven by turbulence is a rapid increase of the aggregate
friction time \tf. Both, theoretical studies\cite{epstein:23} and
experiments\cite{blum:96} show, that for particles, which are small
compared to the mean free path of the gas, \tf is proportional to
their mass--surface ratio. This has the consequence, that the
\tf-evolution is determined by the structure of the growing
aggregates. The increase of the particle mass $m$ with respect to a
typical radius $R$ is a characteristic scaling property of any growth
regime and can be expressed by a fractal dimension \dm: $m(\lambda R)
= \lambda^{\dm} m(R) $.\\

\pict{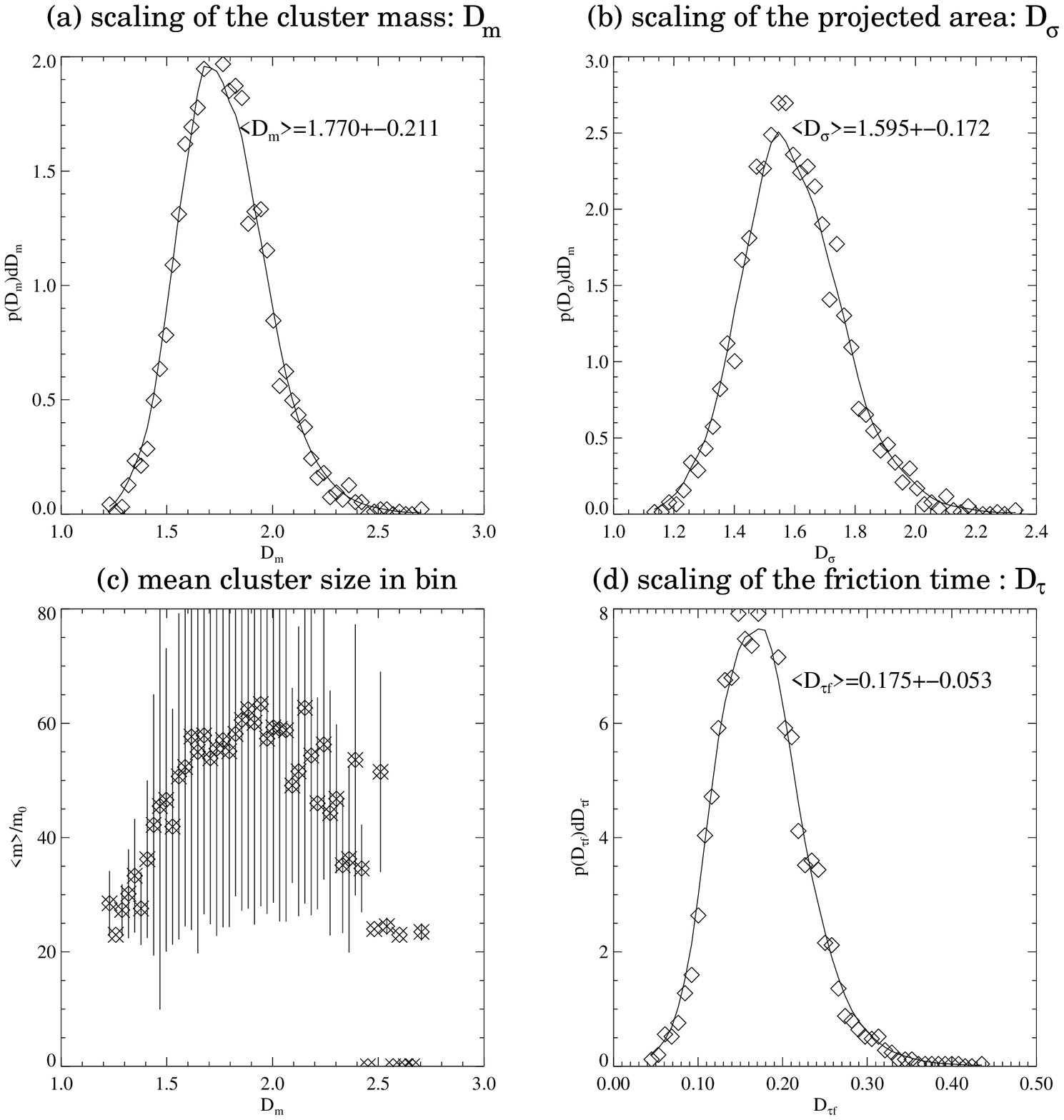}{Normalised distribution of the scaling
  exponents of the cluster mass (Fig. a), projected cluster area (Fig.
  b), and friction time (Fig. d) of all clusters consisting of more
  than 20 single spheres after 355 days of growth. The initial dust
  number density of the monomers was $\nzero=2\cpot{7}\mpot{-3}$. Fig.
  c shows the mean size and size variation of the clusters
  contributing to the bin.}{fig:distr}{0.975}{0.8}{0.075}{0.0}{0.785}
All published models for the dust coagulation assume that there is a
unique value of the fractal dimension \dm describing all aggregates of
the dust ensemble. However, real growth processes produce a wide
variety of different particle structures. In (\figref{fig:distr}a) the
distribution of the fractal dimension of a simulated dust ensemble for
fixed times is shown. We found that the mean fractal dimension
\mean{\dm} is about $1.8$. This value is in good agreement with the
results of Monte-Carlo simulations of the ballistic
Cluster-Cluster-Aggregation\cite{smirnov:90} (BCCA).  In order to
analyse the \tf evolution also the scaling of the projected aggregate
area $\mean{\sigma}$ with the typical radius has to be known. Our
studies show (\figref{fig:distr}b) that the often used assumption of
$\sigma \sim R^2$ implies the wrong picture of the \tf evolution. We
found that $\sigma$ scales approximately as $R^{1.6}$, corresponding
to a friction approximately proportional to the eighth root of the
typical aggregate mass (\figref{fig:distr}c). Although we obtained in
contrast to simplier models a size-dependent friction time our result
still shows that growth solely driven by Brownian motion cannot form
aggregates with friction times comparable to the lifetime of the
smallest turbulent eddies within a million years. However, due to the
broad distribution of the fractal dimension there are some aggregates
growing as compact particles. These particles are of great importance
because they could be possible seed grains for a runaway growth due to
differential sedimentation.
 
\subsection{Evolution of the mass spectrum}

In the previous section we discussed the friction time growth with the
aggregate size. Equally important for the understanding of the
formation of larger aggregates in proto-planetary discs is the
dynamics of the growth. The evolution of the cluster ensemble is
characterised by the number density \ns of clusters of the mass $s$.
For the simulated spectra of the Brownian coagulation we observed a
self-similar evolution. For such growth processes the mass spectrum
can be described by the scaling ansatz\cite{vicsek:84,kolb:84}
\begin{equation}
  \ns (t) \sim t^{-w} s^{-\tau} f(s/t^z),
  \label{ns_crit}
\end{equation}
where $w$, $\tau$, and $z$ are scaling exponents and $f(x)$ is the
scaling function of the growth process. In the asymptotic limit, the
scaling exponents are related by $w=(2-\tau)z$. Furthermore, the mean
cluster mass $S(t) \sim \sum_s^\infty \ns s^2$ evolves asymptotically
as $S(t)\sim t^z$ while the cluster number density $n(t) =
\sum_s^\infty \ns$ scales as $n(t) \sim t^{-z}$ for $\tau < 1$ and
$n(t)\sim t^{-w}$ otherwise\cite{vicsek:84}.
\pict{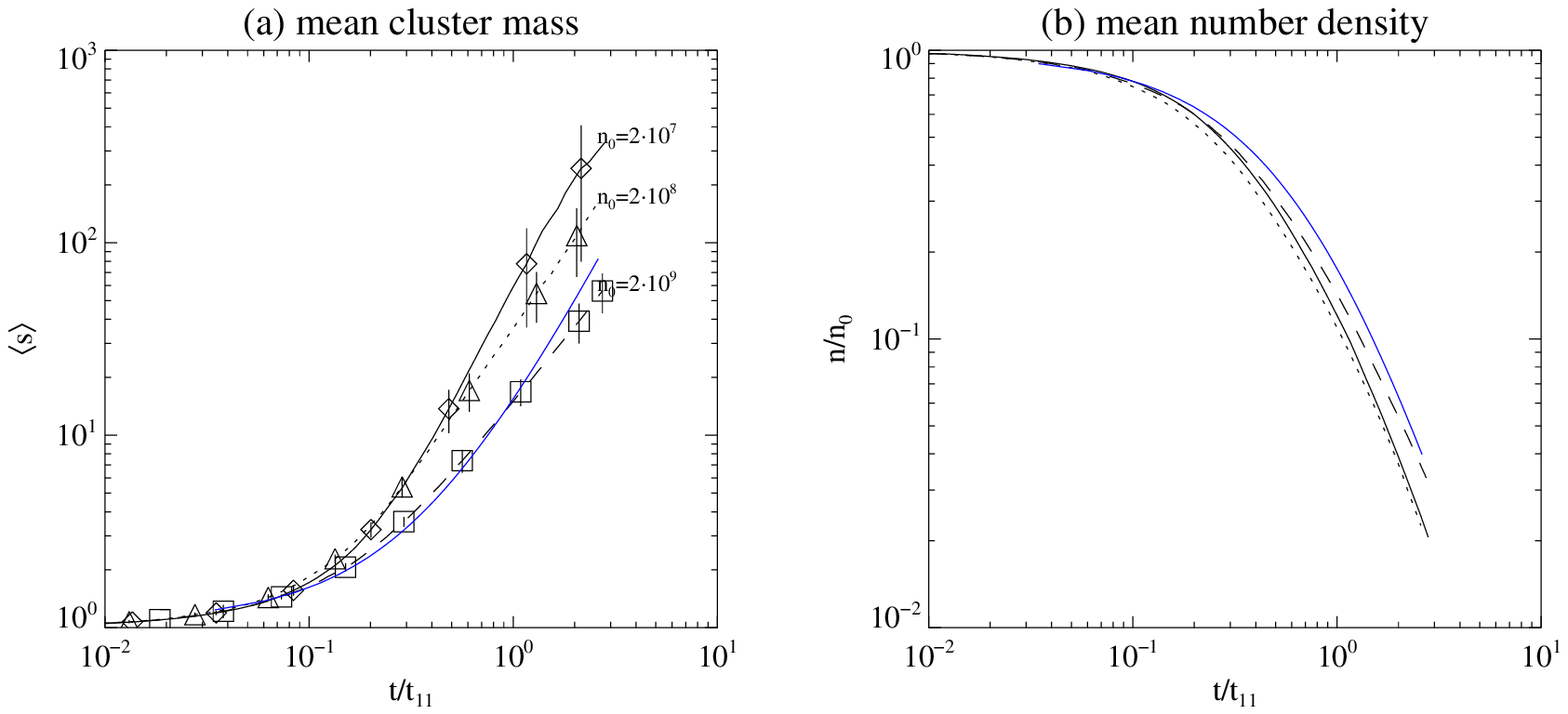} {Normalised evolution of the mean cluster size
  and the cluster number density for initial monomer number densities
  \nzero of $2\cpot{7} \mpot{-3}$ (full line), $2\cpot{8} \mpot{-3}$
  (dotted line), and $2\cpot{9} \mpot{-3}$ (dashed line). The
  dashed-dotted line corresponds to the predictions based on the
  coagulation theory for the ballistic growth of fractals
  \cite{ossenkopf:93}. $t_{11}=\{K_{11}\nzero\}^{-1}$ is the mean
  collision time between monomers.}{fig:exp}
{0.975}{.44}{-0.05}{0.}{1}
>From the averaged spectra we calculated the mean cluster size and the
cluster number density for different initial number densities \nzero
(\figref{fig:exp}). The calculated scaling exponents are related to
each other as predicted by this concept. Consequently, the scaling
ansatz \eqref{ns_crit} is appropriate to describe the dust growth
driven by Brownian motion at low number densities. The scaling
exponents $z$ derived from the asymptotic growth of the mean cluster
mass for different \nzero are smaller than 2. Consequently, the
fastest increase of the mean friction time due to Brownian growth is
$\mean{\tf} \sim t^{1/4}$.  This result supports the point made in the
previous section that the short time scales of dust growth required by
astronomical and geophysical constraints can not be due to coagulation
driven by
Brownian motion.\\

In numerical studies of the cosmic dust growth the aggregation process
is very often described by a rate equation
approach\cite{smoluchowski:17}. As the spatial dependence of the
number densities is neglected, this approach is a mean field
description of the growth. In this framework the scaling exponents
$z$, $w$, and $\tau$ are ruled by the scaling behaviour of the rate
coefficients and do not depend on the initial number density \nzero.
However, as indicated in \figref{fig:exp}, our simulation results show
a significant dependence upon \nzero, which might be due to spatial
fluctuations of the number densities. Our results suggest that using
the Smoluchowski theory should be done with caution
for describing the dust growth in proto-planetary discs.\\

\section{Conclusions}

In the early stages of dust growth the coagulation of dust particles
driven by Brownian motion plays an important role. In order to explain
the short time scales of dust growth as constrained by astronomical
and geophysical studies not only the mass of the dust particles but
also their friction time has to increase quickly. We developed an
N-particle model that allowed us to study coagulation driven by
Brownian motion. In contrast to earlier studies, the dust growth was
investigated at astrophysically relevant small number densities. Under
such conditions the mean particle distance exceeds the typical
aggregate diameter by orders of magnitude, and a collision will be a
rare event. We derived a criterion which allows an efficient detection
of candidates for imminent collisions. The N-particle-method is based
upon an adaptive time step scheme respecting the individual dynamical
states of the aggregates. Its basic concept is to perform on average
constant ``length steps'', instead of using constant time steps. In
order to minimise the influence of the decreasing number of particles
within the simulation box, a new rescaling
method is used throughout the aggregation process.\\

We found that the structural properties of a dust cloud is
characterised more completely by the distribution of the fractal
dimensions of its constituent clusters. The mean of this distribution
of $\dm \sim 1.8$ is in good agreement with the results of Monte-Carlo
simulations of the ballistic Cluster-Cluster Aggregation (BCCA). The
friction time of the aggregates growing due to Brownian coagulation
scales with the
mass as $\mean{\tf} \sim m^{1/10}$.\\

In addition, the dynamical evolution of the dust ensemble has been
investigated.  We found that the mass spectrum of the ensemble evolves
self-similar. For that reason, the mass spectrum of the dust growth
due to Brownian motion at low number densities can be described by a
scaling ansatz. We derived from the simulated spectra that the mean
friction time of the dust ensembles increases with time as $\mean{\tf}
\sim t^{1/4}$. This indicates that the fast dust growth in
proto-planetary is not due to the Brownian coagulation.

%%%%%%%%%%%%%%%%%%%%%%%%%%%%%%%%%%%%%%%%%%%%%%%%%%%%%%%%%%%%%%%%%%%%%%%%%%%%%
%% Acknowledgments
%%%%%%%%%%%%%%%%%%%%%%%%%%%%%%%%%%%%%%%%%%%%%%%%%%%%%%%%%%%%%%%%%%%%%%%%%%%%%
\section*{Acknowledgments}

The authors are indebted to R. Mucha, J. Blum, H. Klahr, and A. Graps 
for helpful and stimulating discussions. This work was made possible by
the support of the DFG to SK (DFG grant He1935 within the special
programme ``Physics of Star-formation'') and SP (DFG-Habilitation
grant). The numerical simulations were partially carried out at the
super-computing centre J{\"u}lich, Germany and we wish to thank for
generous allocation of computing time.

%%%%%%%%%%%%%%%%%%%%%%%%%%%%%%%%%%%%%%%%%%%%%%%%%%%%%%%%%%%%%%%%%%%%%%%%%%%%%
%% Bibliography
%%%%%%%%%%%%%%%%%%%%%%%%%%%%%%%%%%%%%%%%%%%%%%%%%%%%%%%%%%%%%%%%%%%%%%%%%%%%%

\end{document}